\documentclass[twoside,11pt,a4paper]{article}
\usepackage{amssymb}
\usepackage{bm}
\usepackage{amsmath}

\usepackage{color,pstcol}

\newcommand{\ytextmodifartititwohundredsixteenvonestepone}[1]
{#1}

\begin{document}
\normalsize
\sloppy

\begin{center}
{\Large
\textbf{The Landau-Feynman transiently open quantum \ system:
entanglement and density operators}}
~\\
~\\

Alain Deville$^\dag$

$^\dag$
Aix-Marseille Universit\'{e}, CNRS, IM2NP UMR 7334, F-13397 Marseille,
France
~\\
~\\

Yannick Deville$^*$

$^*$
Universit\'{e} de Toulouse, CNRS, CNES, OMP, IRAP (Institut de
Recherche en Astrophysique et Plan\'{e}tologie),
F-31400 Toulouse, France

\ \ \ 
\end{center}

\textbf{Abstract \ }

Users of quantum mechanics, both in physics and in the field of quantum
information, are familiar with the concept of a statistical mixture as
introduced by von Neumann, and with the use of a density operator in that
context. A density operator may also be used in another situation,
introduced by Landau, with a transient coupling between the two parts of a
quantum bipartite system. But more than fifty years after a clarifying work
by Feynman on the subject, a confusion still persists about what we call the
Landau-Feynman situation. This is specifically testified by the development
of controversies around that subject. The aim of this paper is to stress
that, when facing the Landau-Feynman situation, the right concept to be
used is not the one of a mixed state (or statistical mixture) - be it
qualified as proper or improper -, but the one of entanglement.

\vspace{10mm}
\textbf{Keywords \ }

Quantum state tomography; 
bipartite system; transient coupling; density operator; entanglement; 
improper mixture.

\section{Introduction}

2025 has been declared the International Year of Quantum Science and
Technology by the General Assembly of the United Nations. The field of
Quantum Information has been developing for several decades, and keeps
stimulating a reflection about major concepts introduced in the very
beginning of Quantum Mechanics. In this Journal, we recently published a
paper \cite{DevilleA2025} in order to answer a question already identified
by Zeh in 1970 \cite{Zeh1970}. The aim of the present paper is to treat a
situation which is quite distinct from the one considered by Zeh and by our
previous paper, but which is still a source of much confusion.
Considerations about the situation discussed by Zeh in \cite{Zeh1970} and
about the situation treated in this paper are more specifically important in
the context of Quantum State Tomography (QST)\ \cite{NielsenChuang2005} or
in that of standard Quantum Process Tomography (QPT) \cite{NielsenChuang2005}%
, two major topics of Quantum Information Processing.

In our paper \cite{DevilleA2025}, we considered an isolated quantum system $%
\Sigma .$ The word \textit{isolated} may be ambiguous, as \textit{\ an
isolated system may be submitted to forces deriving from a scalar and/or a
vector potential.\ In the absence of any force, the system is said to be free%
}. An isolated system obeys the Schr\"{o}dinger equation, which is
postulated by Quantum Mechanics (QM), and mobilizes the Hamiltonian of the
system, a concept introduced from classical mechanics. An isolated system is
also sometimes said to be closed, and then, if not closed, the system is
said to be open (existence of a coupling Hamiltonian between that system and
its environment).

If the system of interest $\Sigma _{1}$ is not isolated, a simple situation
is the following one: 1) that system $\Sigma _{1}$ is coupled to another
system $\Sigma _{2}$, and the whole system $\Sigma $, composed of these two
parts (bipartite system), is itself isolated, and 2) moreover, one is able
to write explicitly the Hamiltonian of the whole system, as the sum of the
Hamiltonian of each subsystem and of a coupling Hamiltonian.\ In contrast,
if e.g. the system of interest is a spin $1/2$, and if this spin is coupled
to a macroscopic system, one will generally not try to write the whole
Hamiltonian. For instance, if that spin is coupled to a bath, which imposes
it its temperature, and in a situation identified as thermal equilibrium,
the effect of the bath is described with a scalar parameter (the bath
temperature, imposed to the spin).

If the global system $\Sigma $, first prepared in a pure state described
with a ket $\mid \Psi (t_{0})>$ in the Dirac formalism, is thereafter
isolated from its environment (and one should understand that, unless
specifically supposed, forces acting upon it and deriving from a scalar
and/or vector potential are still present), its future behaviour is still
described with a pure state, $\mid \Psi (t)>,$ which obeys the Schr\"{o}%
dinger equation. It may happen that, at $t_{0}$, the state of that isolated
system $\Sigma $ is not pure, but can only be described with a so called
(von Neumann) mixed state or statistical mixture, a collection of (normed)
kets \{$\mid \psi _{i}>,$ $p_{i}$\}, where the randomly drawn ket $\mid \psi
_{i}>$ has the statistical weight $p_{i}$ in the mixture (for any $p_{i},$ $%
0\leq p_{i}$ $\leq 1$ and $\Sigma _{i}p_{i}$ $=1$). The mean value of an
observable $\widehat{O}$ in the mixture, i.e. the quantity $\Sigma
_{i}p_{i}<\psi _{i}\mid \widehat{O}\mid \psi _{i}>$, can be written as $%
Tr\rho \widehat{O},$ where $\rho $ $=$ $\Sigma _{i}p_{i}\mid \psi _{i}><\psi
_{i}\mid $ is the density operator$.$\ As stressed e.g.\ by Peres \cite%
{Peres1995}, considering that $\rho $ contains the whole information present
in the definition of the statistical mixture through the \{$\mid \psi _{i}>,$
$p_{i}$\} collection is not the result of some established property, as
sometimes thought, but the result of a postulate made by von Neumann \cite%
{vonNeumann1932}. \ We discussed the relevance of that postulate in our
recent paper \cite{DevilleA2025}. The present paper is focused \textit{on
another situation} in which a density operator is also introduced, which we
call the Landau-Feynman situation for reasons to be shortly developed.

Thinking that discussions about the use of the density operator (also,
possibly improperly, called the density matrix), or about the pure state
concept are presently out of date and should be moved into the domain of the
history of Science is not consistent with the present reality. For instance,
a 2022 paper by Castellani \cite{Castellani2022} is entitled \textit{All
quantum mixtures are proper}, and, in fact, its content is precisely related
to the Landau-Feynman situation. The concept of an improper mixture was
already present in a 1966 paper by Bernard D'Espagnat \cite{D'Espagnat1966},
and the existence of the Castellani paper does testify that, some sixty
years later, the thinking about that subject is not stabilized yet. And,
specifically in the context of QST, which aims at estimating quantum states,
and in that of standard QPT, with its use of QST, it would be useful to
fully appreciate the content of that Castellani paper. In Section \ref%
{SectionD'EspagnatCastellani} of the present paper, the idea behind the
improper mixture concept introduced by d'Espagnat, which recently led to the
2022 paper by Castellani, is presented and discussed.

When speaking of a von Neumann mixed state, one generally refers to the
already cited 1932 canonical book by von Neumann \cite{vonNeumann1932}. In
fact, a density operator $\rho $ had already been introduced in two
different physical contexts, by von Neumann \cite{vonNeumann1927} (the
situation he studied in more detail in his 1932 book) and by Landau \cite%
{Landau1927}, in the case of permanently coupled systems.\ None of these
contexts corresponds to what we call the Landau-Feynman situation, which
appeared in volume III of the course by Landau and Lifschitz \cite%
{Landau1965}, and which is presented in Section \ref{SectionLandau}. The
analysis made by Feynman in his 1972 book focused on quantum 
\ytextmodifartititwohundredsixteenvonestepone{statistical mechanics}
\cite%
{Feynman1972} is detailed in Section \ref{SectionFeynman}.\ It allows one to
appreciate the role of entanglement, and the conditions to be fulfilled when
a density operator is introduced in that situation.\ These results,
discussed in Section \ref{SectionDiscussion}, show that both the claim by
Castellani and the concept of an improper mixture should be rejected. A
conclusion is given in Section \ref{SectionConclusion}. A simple instance of
the Landau-Feynman situation, implying two distinguishable spins 1/2
transiently coupled with a Heisenberg cylindrical exchange coupling, is
examined in Appendix 1.\ In \ Appendix 2, considering the Landau-Feynman
situation, it is shown that, whereas it it possible to introduce two density
operators
$\rho _{1}$ and $\rho _{2}$ through partial tracing, it would
generally be wrong to claim that the state of the whole system $\Sigma $ is $%
\rho _{1}$ $\otimes $ $\rho _{2}$.

\section{From d'Espagnat to Castellani\label{SectionD'EspagnatCastellani}}

In a 1966 paper \cite{D'Espagnat1966}, D'Espagnat \ considered an implicitly
isolated system composed of two parts, each one being initially in a pure
state.\ The two subsystems then interacted, and were then again uncoupled. 
\textbf{This situation is precisely the one which we call the Landau-Feynman
situation}. D'Espagnat observed that, after the disappearance of that
interaction, the whole system was still in a pure state, and that one could
be interested in only one of the subsystems. He then wrote: "\textit{after
the interaction has taken place, one immediately sees that, in general,
neither of them is a pure case. One then usually says that they are a
mixture".\ }D'Espagnat then proposed to say that they are in "an improper
mixture", and that a von Neumann mixture is "a proper mixture".

In this paper, we stress that D'Espagnat did identify a problem, but that
introducing the expression "improper mixture" kept an ambiguity, and
therefore did not solve that problem.\ Its persistence is testified by the
existence of the 2022 paper \cite{Castellani2022} by Castellani. From the
beginning of his paper, Castellani:

\begin{itemize}
\item assumes "that the state of a system is completely described by its
density operator", which in fact is just a postulate (cf. \cite{DevilleA2025}
and its references),

\item considers "the state of a subsystem in a composite system to be
completely described by its reduced density matrix". As detailed in Sections %
\ref{SectionFeynman} and \ref{SectionDiscussion}, this claim is wrong.
\end{itemize}

The Castellani paper, which expresses a disagreement with {an article
published by D'Espagnat in 2001 }\cite{D'Espagnat2001} (itself a reply to a
2001 paper from K.A. Kirkpatrick \cite{Kirkpatrick2001}), ignores the
content of the 1970 Zeh paper \cite{Zeh1970}. That paper by Castellani
refers to the 1932 book by von Neumann \cite{vonNeumann1932}, in which
anyway what we call the Landau-Feynman situation was not discussed. The
approach taken by Castellani in \cite{Castellani2022} moreover ignores the
content of the 1972 book \cite{Feynman1972} by Feynman on the subject. {The
whole approach followed by Castellani in his paper \cite{Castellani2022}\ }%
imposes us to detail the concise {treatment given by Feynman in \cite%
{Feynman1972}, which will be done in Section \ref{SectionFeynman}.}

\section{Landau and the use of a density operator\label{SectionLandau}}

Texts briefly considering the historical introduction of the density
operator generally claim that both Landau and von Neumann introduced the
concept in 1927. It is true that, in a 1927 paper \cite{Landau1927}, Landau
(born in 1908), in the first section, entitled "Coupled systems in wave
mechanics"\textit{, }wrote: "A system cannot be uniquely defined in wave
mechanics; we always have a probability ensemble (statistical treatment). If
the system is coupled with another, there is a double uncertainty in its
behaviour\textit{". }Then, considering an observable attached to the first
subsystem, \textit{in the presence of such a coupling, }and using the
formalism of the wave function, he introduced an operator through an
integration over the variables of the second subsystem, which corresponds to
a \textit{Partial Trace} procedure. And, unambiguously, the title of his
paper contains the word damping (in \ its English version \cite{Landau1927})
associated with an irreversible process if, in the instance treated by
Landau, one focuses on the electrons. Such a permanent (as opposed to a
transient) coupling is \textit{not }what, in this paper, we call the
Landau-Feynman situation.

Roughly thirty years later (1956, Russian version, and page 38 of the
English 1965 version \cite{Landau1965}), in volume III of their course,
devoted to non-relativistic quantum mechanics, Landau and Lifshitz first
supposed that a\textit{\ "}closed system as a whole is in some state
described by a wave function $\Psi (q,x),$\ where $x$ denotes the set of
coordinates of the system of interest, and $q$ the remaining coordinates of
the whole system\textit{". }Integrating over the $q$ variables -which
corresponds to introducing a partial trace, they introduced an operator 
\textit{which they again called a density matrix} (this should today
correspond to the concepts of a master equation and Lindblad operator, or
generalized density operator, cf.\ our brief comment in Section \ref%
{SectionDiscussion}).\textbf{\ Then, in a second step only, they "suppose
that the system" (of interest, corresponding to }$x$\textbf{) "is closed, or
became so at some time".\ That second situation is the one which we call the
Landau-Feynman situation}. And it is the situation considered by d'Espagnat
(and by K.A. Kirkpatrick or Castellani, cf. Section \ref%
{SectionEspagnatCastellani}\label{SectionEspagnatCastellani}).

As explained in Chapter{\ }2 of his \textit{Statistical Mechanics} \cite%
{Feynman1972}, and developed in Section \ref{SectionFeynman}, Feynman
suppressed the possible confusion resulting from the use of the expression
density or statistical operator by both von Neumann and Landau under
different assumptions, and more specifically with the possible existence of
a transient coupling of the system of interest with a second system in the
Landau and Lifschitz course\textit{.}

\section{The treatment of the Landau case by Feynman\label{SectionFeynman}}

Chapter 2 of the quite synthetic 1972 book by Feynman, \textit{Statistical
Mechanics }\cite{Feynman1972}\textit{,} is entitled Density matrices. That
book is rarely cited, and, instead of just mentioning its existence, we
hereafter detail the argumentation developed by Feynman on the situation
introduced by Landau (Section \ref{SectionLandau}).

Feynman divides the universe into two parts - the system of interest (we
call it $\Sigma _{1}$) and the rest of the universe - and we will consider
that the rest of the universe is just the environment of the system of
interest at our chose time scale, and call it $\Sigma _{2}$. He then
introduces a complete set of orthonormal kets, \{$\mid \varphi _{i}>$\}, for 
$\Sigma _{1},$ and a complete set of orthonormal kets, \{$\mid \theta _{j}>$%
\}, for $\Sigma _{2}.$\ The most general (normed) pure state for the whole
system can be written as:%
\begin{equation}
\mid \psi >=\Sigma _{i,j}c_{ij}\mid \varphi _{i}>\mid \theta _{j}>.
\label{PsiBipartite}
\end{equation}%
Feynman introduces an observable acting on the system of interest only, i.e.
of the form $A\otimes I_{2}$ ($I_{2}$: the unit operator in the state space
of $\Sigma _{2}$). He considers the mean value of this observable in state $%
\mid \psi >,$ i.e. the quantity:%
\begin{eqnarray}
<\psi \mid A\otimes I_{2}\mid \psi >
&
=
&
\Sigma _{iji^{\prime }j^{\prime
}}c_{ij}^{\ast }c_{i^{\prime }j^{\prime }}<\theta _{j}\mid <\varphi _{i}\mid
A\otimes I_{2}\mid \varphi _{i^{\prime }}>\mid \theta _{j^{\prime }}> 
\nonumber
\\
&=&\Sigma _{iji^{\prime }}c_{ij}^{\ast }c_{i^{\prime }j}<\varphi _{i}\mid
A\mid \varphi _{i^{\prime }}>,
\end{eqnarray}%
which can be written as $<\psi \mid A\otimes I_{2}\mid \psi >=\Sigma
_{ii^{\prime }}<\varphi _{i}\mid A\mid \varphi _{i^{\prime }}>\rho
_{1,i^{\prime }i}$ with $\rho _{1,i^{\prime }i}=\Sigma _{j}c_{ij}^{\ast
}c_{i^{\prime }j}$. Feynman then defines an operator acting in the Hilbert
space of $\Sigma _{1}$, which we will note $\rho _{1}$ (Feynman notes it as $%
\rho $), and such that, for any $i^{\prime },$ $i:$ $\rho _{1,i^{\prime
}i}=<\varphi _{i^{\prime }}\mid \rho _{1}\mid \varphi _{i}>.$

Manifestly, for any $i^{\prime },i$: $\Sigma _{j}c_{ij}^{\ast }c_{i^{\prime
}j}=(\Sigma _{j}c_{i^{\prime }j}^{\ast }c_{ij})^{\ast }$ i.e. $\rho
_{1,i^{\prime }i}=\rho _{1\text{,}ii^{\prime }}^{\ast }$ , which means that $%
\rho _{1}$ is Hermitian. Therefore $\rho _{1}$ does possess a diagonal form,
and its eigenvalues are real. Moreover, since a Trace does not depend on the
chosen basis, and since $\mid \psi >$ is normed, this trace is equal to one:%
\begin{equation}
Tr\rho _{1}=\Sigma _{i}\rho _{1,ii}=\Sigma _{ij}\mid c_{ij}\mid ^{2}=1.
\label{TraceDeRho1}
\end{equation}%
Feynman then shows that all the eigenvalues of $\rho _{1}$ are non-negative.
Detailing his approach, we introduce the operator $A\otimes I_{2},$ with $%
A=\mid i^{\prime }><i^{\prime }\mid $, $\mid i^{\prime }>$ being an
eigenstate of $\rho _{1},$ with eigenvalue $w_{i^{\prime }}$: $\rho
_{1}=\Sigma _{i}w_{i}\mid i><i\mid .$\ We then calculate the mean value of $%
A\otimes I_{2}$ \ \ in state $\mid \psi >:$%
\begin{eqnarray}
<\Psi \mid \mid i^{\prime }><i^{\prime }\mid \otimes I_{2}\mid \Psi >
&=&<\Psi \mid i^{\prime }><i^{\prime }\mid \otimes \Sigma _{j}\mid \theta
_{j}><\theta _{j}\mid \Psi > \\
&=&\Sigma _{j}<\Psi \mid i^{\prime }>\mid \theta _{j}><i^{\prime }\mid
<\theta _{j}\mid \Psi > \\
&=&\Sigma _{j}\mid <i^{\prime }\mid <\theta _{j}\mid \Psi >\mid ^{2}.
\end{eqnarray}%
This mean value $<\Psi \mid \mid i^{\prime }><i^{\prime }\mid \otimes
I_{2}\mid \Psi >$ is also equal to:%
\begin{eqnarray}
Tr\{\rho _{1} \mid i^{\prime }><i^{\prime }\mid \}
&
=
&
Tr\{\Sigma
_{i}w_{i}\mid i><i\mid \mid i^{\prime }><i^{\prime }\mid \} \\
&=&Tr\{
w_{i^{\prime }}\mid i^{\prime }><i^{\prime }\mid \}
\\
&
=
&
w_{i^{\prime
}}.
\end{eqnarray}%
Therefore: $w_{i^{\prime }}=\Sigma _{j}\mid <i^{\prime }\mid <\theta
_{j}\mid \Psi >\mid ^{2}:$ any eigenvalue of $\rho _{1}$ is non-negative.

\ \ 

In the situation analyzed by von Neumann, with a statistical mixture $\{\mid
\Psi _{i}>,$ $p_{i}\}$ of normed kets $\mid \Psi _{i}>,$ examined in our
paper \cite{DevilleA2025}, the density operator $\rho =\Sigma _{i}p_{i}\mid
\Psi _{i}><\Psi _{i}\mid $ was a formal tool allowing one to get an
expresssion of the mean value of an observable $\hat{O}$ in the presence of
that statistical mixture as a Trace: 
$Tr\rho \hat{O}$. 
In the present
situation, a density operator $\rho _{1}$ acting in the state space of $%
\Sigma _{1}$ has been introduced, allowing one to write that the mean value
of $A$, an observable of $\Sigma _{1}$, is equal to $Tr\rho _{1}A.$\ One may
now introduce the following density operator: $\rho _{w}$ (w: whole):%
\begin{equation}
\rho _{w}=\mid \psi ><\psi \mid ,
\end{equation}%
describing the whole system in state $\mid \psi >$.\ The introduction of
both $\rho _{w}$ and $\rho _{1}$ allows one to describe the contribution of
the collection of the normed states $\mid \varphi _{i^{\prime }}>\mid \theta
_{j^{\prime }}>$ for a given $i^{\prime }$ in a compact way: one can verify
that $\rho _{1}$ can be interpreted as the following partial Trace (over the
state space of $\Sigma _{2}$):%
\begin{equation}
\rho _{1}=Tr_{\Sigma _{2}}\rho _{w}.
\end{equation}%
If, at some time $t_{0},$ the system and its environment are both prepared
in a pure state, and if, then, during a time interval $\tau ,$ a coupling
exists between the system and its environment, at the end of this time
interval, here taken as the origin of time ($t_{0}$ being therefore
negative), the whole system is still in a pure state $\mid \psi (0)>$,
obeying the Schr\"{o}dinger equation, but, except accidently, $\mid \psi
(0)> $ is not a product state. If, at $t=0,$ one is interested in $A$ acting
in the state space of $\Sigma _{1}$ only, one can introduce the density\
operator $\rho _{w}=\mid \psi (0)><\psi (0\mid $ describing the state of the
whole system $\Sigma $, and then, using a partial Trace, obtain the mean
value $<\psi (0)\mid A\mid \psi (0)>.$\ One may be interested in the mean
value of $A$ at a time $t>0$, i.e. after the disappearance of the transient
coupling of $\Sigma _{1}$ with $\Sigma _{2}$ (but within, of course, our
time scale). Feynman introduces the eigenstates $\mid E_{n}>$ of $H_{1},$
the Hamiltonian of\ $\Sigma _{1}:$ $H_{1}\mid E_{n}>=E_{n}\mid E_{n}>$.\
Then, if $f(H_{1})$ is a function of $H_{1},$ $f(H_{1})\mid
E_{n}>=f(E_{n})\mid E_{n}>$ (definition of a function of an operator). He
then first writes $\rho _{1}$ (our own notations are kept here) as $\rho
_{1}=\Sigma _{i}w_{i}\mid i(0)><i(0)\mid $, ($\mid i(0)>$ was previously
written as $\mid i>$), and then $\rho _{1}(t)$ as $\rho _{1}(t)=\Sigma
_{i}w_{i}\mid i(t)><i(t)\mid $. Detailing the argumentation from Feynman 
\cite{Feynman1972}, one first observes that, for $t\geq 0:$%
\begin{eqnarray}
\mid i(t)>
&
=
&
\Sigma _{n}\mid E_{n}><E_{n}\mid \mid i(t)> \\
&=&\Sigma _{n}\mid E_{n}>e^{-iE_{n}t/\hslash }<E_{n}\mid \mid i(0)> \\
&=&e^{-iH_{1}t/\hslash }\mid i(0)>.
\end{eqnarray}%
Since $\rho _{1}(t)=\Sigma _{i}w_{i}\mid i(t)><i(t)\mid $, one gets:%
\begin{eqnarray}
\rho _{1}(t) &=&\Sigma _{i}w_{i}e^{-iH_{1}t/\hslash }\mid i(0)><i(0)\mid
e^{iH_{1}t/\hslash } \\
&=&e^{-iH_{1}t/\hslash }\rho _{1}(0)e^{iH_{1}t/\hslash }.
\end{eqnarray}%
Then deriving with respect to time, one gets, for $t\geq 0:$ 
\begin{equation}
i\hslash \frac{d\rho _{1}}{dt}=[H_{1},\rho _{1}],
\end{equation}%
(Feynman used units with $\hslash =1$), \ an equation also obeyed in the
presence of a von Neumann mixture, and usually called the Liouville - von
Neumann equation.

\section{Discussion\label{SectionDiscussion}}

In Section \ref{SectionFeynman} it was explained that if a system $\Sigma $
composed of two parts $\Sigma _{1}$ and $\Sigma _{2}$ is isolated and
described by a density operator $\rho _{w}$ (w: whole), one may introduce
either $\rho _{1}=Tr_{\Sigma _{2}}\rho _{w}$ or $\rho _{2}=Tr_{\Sigma
_{1}}\rho _{w},$ a formal tool helpful in the calculation of the mean value
of an observable attached to one subsystem only, once a possible coupling
between the two subsystems has disappeared. As detailed in Appendix 2, it
would generally be quite wrong to think that the whole system is then
described by $\rho _{1}\otimes \rho _{2},$ generally describing a
statistical mixture, whereas one knows that the whole system is in a pure
state $\mid \Psi (t)>.$ Already in 1939, F. London and Bauer wrote (English
translation, page 248 in \cite{LondonBauer1939}): "\textit{While the
combined system I+II, which we suppose isolated from the rest of the world,
is and remains in a pure state, we see that during the interaction systems I
and II individually transform themselves from pure cases into mixtures.\
This is a rather strange result". }In fact, the concept to be used in that
situation is not the one of a mixture, but the concept of \textit{%
entanglement:~}with the notations used by F. London and Bauer, I and II are
initially in a product state, and, as a consequence of their interaction,
the state of the whole system becomes an entangled (pure) state, and then it
is meaningless to try and speak of the state of either subsystem I or
subsystem II. The word entanglement had been introduced four years earlier
by Schr\"{o}dinger in the context of the EPR discussion.\ That problem
mentioned by F.\ London and Bauer was not treated in the 1955 canonical book 
\textit{Quantum Mechanics }from Schiff \cite{Schiff1955}. Since the
appearance of the 1972 book on 
statistical mechanics from Feynman, the
problem was presented neither in the book \textit{Quantum Theory: Concepts
and Methods}\ by Peres \cite{Peres1995} nor in the 2013 \textit{Lectures on
Quantum Mechanics }by Steven Weinberg \cite{Weinberg2013}.\ It was discussed
by Le Bellac in his 2006\ book \cite{LeBellac2006}.\ But Le Bellac adopted
the point of view introduced by D'Espagnat, with a distinction between
so-called proper and improper mixtures, instead of identifying the presence
of a pure, entangled, state.

The reader interested in the situation found when a coupling Hamiltonian
does persist between the two parts of an isolated bipartite sytem, i.e. the
situation when each subsytem is said to be open, may consult the book 
\textit{The Theory of open quantum systems, }by Breuer and Petruccione \cite%
{Breuer2002}, and identify the importance of the so-called master (or
Lindblad) equation in that situation.

\section{Conclusion\label{SectionConclusion}}

When the two parts of an isolated bipartite system are initially in a pure
state, and are then momentarily coupled to each other, one may focus on one
of these subsystems once the internal coupling has disappeared. If one wants
to calculate the mean value of some observable attached to that subsystem,
one may introduce a density operator, through partial tracing over the
second subsystem. A 2022 paper and a 2006 book, both cited in this paper,
show that the meaning of that density
\ytextmodifartititwohundredsixteenvonestepone{operator}
is still under debate. The aim
of the present paper was to stress once for all that it would be wrong to
try and \textit{interpret} the introduction of this so-called reduced
density 
\ytextmodifartititwohundredsixteenvonestepone{operator}
through a reference to some mixed state, be it either proper
or improper. The right concept in this context is the one of entanglement:
once the transient internal coupling has disappeared, the bipartite system
keeps in an entangled pure state (except when, either it exceptionnally
keeps in a pure product state, or, at some specific times, it accidentally
becomes unentangled).

~\\
{\bf
Appendix 1%
.
Determination of a mean value with the reduced density
operator \label{SectionAppendixOne}}

The following simple example of the treatment of the Landau-Feynman
situation mobilizes two distinguishable 
\ytextmodifartititwohundredsixteenvonestepone{spins}
$1/2,$ $\overset{\rightarrow }{%
s_{1}}$ and $\overset{\rightarrow }{s_{2}},$ initially prepared in the
product pure state $\mid $ $\Psi _{1}>\mid \Psi _{2}>$ (a short writing for$%
\mid 1,$ $\Psi _{1}>\otimes \mid 2,$ $\Psi _{2}>$ ), \ with:%
\begin{eqnarray}
&\mid &\Psi _{1}>=\alpha _{1}\mid +>+\beta _{1}\mid -> \\
&\mid &\Psi _{2}>=\mid +>.
\end{eqnarray}%
From this time, and during a time interval with a duration equal to $\tau $,
these spins are submitted both to a Zeeman coupling (static magnetic field $%
\overrightarrow{B}$ along $Oz$), and to a Heisenberg exchange coupling with
cylindrical symmetry, leading to the following Hamiltonian:%
\begin{equation}
\mathcal{H=}G(s_{1z}+s_{2z})B-2J_{xy}(s_{1x}s_{2x}+s_{1y}s_{2y}).
\end{equation}%
The expression of the pure but (perhaps except, accidentally, at some
specific times) entangled state at the end of this time interval $\tau $ can
be directly established (or deduced from\ \cite{DevilleY2012}, as $\mathcal{H%
}$ is identical with the one in Eq. (4) of \cite{DevilleY2012}, if one takes 
$J_{z}=0,$ $\mid \Psi _{1}>$ , $\mid \Psi _{2}>$ then corresponding to $\mid
\Psi _{1}(t_{0})>$ and $\mid \Psi _{2}(t_{0})>$ respectively, with $\beta
_{2}=0$ and $\alpha _{2}=1$ in Eq. (6) of \cite{DevilleY2012}).\ We
presently take the end of this time interval with duration $\tau $ (i.e. the
time when this transient coupling disappears: $t_{0}$ is therefore negative)
as the origin of time. In order to write the expression of this entangled
pure state $\mid \Psi (0)>$ at the end of this transient moment, it is
useful to introduce four quantities corresponding to possible energy
differencies (cf.\ Eq. (64)-(67) in \cite{DevilleY2012}), and three of them
are mobilized here:%
\begin{equation}
\omega _{1,1}=GB/\hslash \text{, \ \ \ }\omega _{1,0}=-J_{xy}/\hslash
=-\omega _{0,0}.
\end{equation}%
From a direct calculation (or from Eq. (68) of \cite{DevilleY2012}), one
gets:%
\begin{equation}
\mid \Psi (0)>=a\mid ++>+b\mid +->+c\mid -+>,
\end{equation}%
with:%
\begin{equation}
a=\alpha _{1}e^{-i\omega _{1,1}\tau }\text{, \ \ }b=-i\beta _{1}\sin \omega
_{1,0}\tau ,\text{ \ }c=\beta _{1}\cos \omega _{1,0}\tau .
\end{equation}%
The pure state $\mid \Psi (0)>$\ at the end of the transient Heisenberg
coupling is therefore entangled, except if the duration $\tau $ of that
coupling is such that either $\sin \omega _{1,0}\tau $ or $\cos \omega
_{1,0}\tau $ is equal to $0.$

If one decides to calculate the mean value of e.g. $\sigma _{1x},$ $\sigma
_{1y}$ or $\sigma _{1z}$ (the Pauli operators for spin $N%
{{}^\circ}%
$ $1$, acting within the state space of $\Sigma _{1}$) at the end of this
transient coupling, one may first calculate the expression of $\rho
_{1}=Tr_{\Sigma _{2}}\mid \Psi (0)><\Psi (0)\mid $, i.e. the quantity%
\begin{equation}
\rho _{1}=<2+\mid \mid \Psi (0)><\Psi (0)\mid \mid 2+>+<2-\mid \mid \Psi
(0)><\Psi (0)\mid \mid 2->.
\end{equation}%
The first term in this expression has the following contribution to $\rho
_{1}:$%
\begin{equation}
(a\mid +>+c\mid ->)(a^{\ast }<+\mid +c^{\ast }<-\mid ),
\end{equation}%
which, when developed, leads to two projectors and two dyads (acting in the
state space of $\Sigma _{1}$):%
\begin{equation}
\mid a\mid ^{2}\mid +><+\mid +\mid c^{2}\mid -><-\mid +ac^{\ast }\mid
+><-\mid +ca^{\ast }\mid -><+\mid .
\end{equation}%
The second term has the following contribution to $\rho _{1}:$%
\begin{equation}
\mid b\mid ^{2}\mid +><+\mid .
\end{equation}%
Finally:%
\begin{equation}
\rho _{1}=(\mid a\mid ^{2}+\mid b\mid ^{2})\mid +><+\mid +\mid c^{2}\mid
-><-\mid +ac^{\ast }\mid +><-\mid +ca^{\ast }\mid -><+\mid ,
\end{equation}%
where the projectors and dyads refer to spin $1.$\ One immediately verifies
that the Trace of $\rho _{1}$ is equal to $1.$

The expression of $\rho _{1}$ is hereafter used in the calculation of the
mean value of $\sigma _{1x}\ $in state $\mid \Psi (0)>$%
\begin{equation}
<\Psi (0)\mid \sigma _{1x}\ \mid \Psi (0)>=Tr\{\rho _{1}\sigma _{1x}\}.
\end{equation}%
The projectors present in $\rho _{1}$ obviously do not contribute to that
mean value, which therefore reduces to:

\begin{center}
\begin{tabular}{l}
$<+\mid (ac^{\ast }\mid +><-\mid +ca^{\ast }\mid -><+\mid )\sigma _{1x}\mid
+>+$%
\end{tabular}%
\begin{equation}
<-\mid (ac^{\ast }\mid +><-\mid +ca^{\ast }\mid -><+\mid )\sigma _{1x}\mid
->,
\end{equation}
\end{center}
and finally:%
\begin{equation}
<\Psi (0)\mid \sigma _{1x}\ \mid \Psi (0)>=ac^{\ast }+ca^{\ast }.
\end{equation}%
One may choose instead to calcultate directly that mean value, i.e the value
of:%
\begin{equation}
(a^{\ast }<++\mid +b^{\ast }<+-\mid +c^{\ast }<-+\mid )\sigma _{1x}(a\mid
++>+b\mid +->+c\mid -+>),
\end{equation}%
where $\sigma _{1x}$ is a short writing for $\sigma _{1x}\otimes I_{2}.$\
The $a^{\ast }<++\mid $ bra introduces one contribution to the mean value,
and the same is true for the bra $c^{\ast }<-+\mid .$ The $b^{\ast }<+-\mid $
bra does not contribute to the mean value, and one finally gets the same
result.

The direct calculation of that mean value of $\sigma _{1x}$ is manifestly
simpler than its calculation with $\rho _{1}$.\ But this paper is not aimed
at finding the easiest way for calculating mean values, but at identifying
the content of the Landau-Feynman situation, and of the introduction of a
density operator through partial tracing in that situation.

~\\
{\bf
Appendix 2.\ A wrong use of partial traces\label{SectionAppendixTwo}%
}

Instead of considering $\Sigma _{1}$ as the system of interest, and then
defining $\rho _{1}=Tr_{\Sigma _{2}}\rho \,,$ someone may choose $\Sigma
_{2} $ as his system of interest, and then define $\rho _{2}=Tr_{\Sigma
_{1}}\rho .$\ In that paper, it has been stressed that, when trying to speak
of the state of the subsystems $\Sigma _{1}$ or $\Sigma _{2}$ in the
Landau-Feynman situation, the right concept to use is that of entanglement,
not that of mixture, be it 
\ytextmodifartititwohundredsixteenvonestepone{proper or improper.}
Someone who keeps speaking of a
mixture may not only \textit{wrongly} consider that $\Sigma _{1}$ and $%
\Sigma _{2}$ are in stastistical mixtures $\rho _{1}$ and $\rho _{2}$
respectively, but then add that the whole system $\Sigma $ is in a state
described as $\rho _{1}$ $\otimes \rho _{2}$.\ If it is objected that it is
known that $\Sigma $ is in a pure state, he may reply that nothing prevents $%
\rho _{1}$ $\otimes \rho _{2}$ from being a projector, associated 
\ytextmodifartititwohundredsixteenvonestepone{with}
a pure
state. In this Appendix, we discuss the relevancy of such a claim.

We first momentarily examine the quite specific following situation: two
distinguishable spins $1/2,$ spin $1$ and spin $2$ which, at time $t_{0},$
(with $t_{0}<0$) are prepared in the product state $\mid 1+>\otimes \mid 2->$
($\mid +>$ being the usual notation for the eigenstate of $s_{z}$ for the
eigenvalue equal to $1/2$ in reduced units). From $t_{0}$ to time $t=0,$ the
spins are submitted to an internal coupling equal to $2J_{zz}s_{1z}s_{2z.}$\
One is interested in the behaviour of the spin pair for $t\geq 0,$ i.e. once
the internal coupling has disappeared.\ The answer is obvious: because of
the form of both the initial state of the spin pair and the transient
internal coupling, each spin lives a life of its own: no entanglement
appears.

We just introduced this situation in order to stress that in this Appendix
we are not interested in such situations.\ Just in contrast, \ having first
written $\rho _{1}=\Sigma _{i}\lambda _{i}\mid i>
\ytextmodifartititwohundredsixteenvonestepone{<i\mid}
,$ where the collection $\
\{\mid i>\}$ is an eigenbasis of normed kets of $\rho _{1},$ with $0\leq
\lambda _{i}\leq 1,$ $\Sigma _{i}\lambda _{i}=1$ (properties of the density
operator $\rho _{1}$), we consider situations where at least two eigenvalues 
$\lambda _{i}$ are not equal to zero (one discards any time when,
accidently, the state of the whole system is unentangled). One then writes $%
\rho _{2}=\Sigma _{j}\mu _{j}\mid j>
\ytextmodifartititwohundredsixteenvonestepone{<j\mid}
,$ with similar definitions and
properties, this time related to $\Sigma _{2}.$

It is now possible to write $\rho _{1}$ $\otimes \rho _{2}$ as:%
\begin{equation}
\rho _{1}\otimes \rho _{2}=\Sigma _{i,j}\lambda _{i}\mu _{j}\mid i>\mid j>
\ytextmodifartititwohundredsixteenvonestepone{<i\mid <j\mid .}
\end{equation}%
From what is known about the $\lambda _{i}$ and $\mu _{j}$, it is impossible
to reduce this double sum to a single projector, which would 
require
all the $%
\lambda _{i}\mu _{j}$ products to be equal to zero, except one of them,
equal to one.

$\rho _{1}\otimes \rho _{2}$ therefore necessary describes a mixed state
(except perhaps, accidentally, at some specific times), and can't describe
the 
\ytextmodifartititwohundredsixteenvonestepone{pure}
state of the whole system $\Sigma $.

\end{document}